\documentclass[12pt]{article}

\usepackage[pdftex]{graphicx}
\usepackage[T1]{fontenc}
\usepackage{amsthm,amsmath,amssymb}
\allowdisplaybreaks

\usepackage[usenames,dvipsnames]{color}
\usepackage[dvipsnames]{xcolor}
\usepackage[pdftex,colorlinks=true,unicode,bookmarksnumbered=true,hypertexnames=false]{hyperref}
\hypersetup{
  colorlinks,
  citecolor=blue!80!black,
  linkcolor=red!80!black,
  urlcolor=blue!80!black}

\usepackage[doublespacing]{setspace}
\usepackage{textcomp}
\usepackage{indentfirst}
\usepackage{natbib}
\usepackage{listings}
\usepackage{upquote}
\usepackage{pdflscape}
\usepackage{enumerate}
\usepackage{verbatim}
\usepackage{bbm}
\usepackage{float}
\usepackage{multirow}
\usepackage{authblk}
\usepackage[toc,page]{appendix}
\usepackage{array,longtable}
\usepackage{booktabs}
\usepackage{mathtools}
\usepackage{geometry}
\DeclareMathOperator*{\argmin}{argmin}

\numberwithin{equation}{section}

\theoremstyle{definition}

\renewenvironment{abstract}
 {\small
  \onehalfspacing
  \hyphenpenalty=10000
  \exhyphenpenalty=10000
  \emergencystretch=2em
  \begin{center}
  \bfseries \abstractname\vspace{-.5em}\vspace{0pt}
  \end{center}
  \list{}{%
    \setlength{\leftmargin}{5mm}%
    \setlength{\rightmargin}{\leftmargin}%
    \listparindent 1.5em
    \itemindent    \listparindent
  }%
  \item\relax}
 {\endlist}

\newcommand{\keywords}[1]{\textbf{Keywords:} #1}
\newcommand{\jel}[1]{\textbf{JEL codes:} #1}
\newcommand{\code}[1]{\texttt{#1}}
\newcommand{\tablenote}{\emph{Notes:}}

\begin{document}

\title{Estimating Semiparametric and Nonparametric Fixed Effects Panel Data Models with \code{mgcv}}

\author{Ivan Korolev\thanks{Department of Economics, Binghamton University. E-mail: \mbox{\href{mailto:ikorolev@binghamton.edu}{ikorolev@binghamton.edu}}. Website: \url{https://sites.google.com/view/ivan-korolev/home}}}

\date{\today}

\maketitle

\begin{abstract}
This paper provides a practical guide to estimating semiparametric and nonparametric fixed-effects panel data models using the \code{mgcv} package in R. The focus is implementation: handling fixed effects with unit indicators, first differencing, or penalized unit effects; specifying smooth terms; and conducting cluster-robust inference. Monte Carlo experiments compare \code{mgcv::bam} estimators with linear and fixed-series spline estimators. Simulations suggest that penalized splines adapt to unknown smoothness and estimate functions accurately in the designs studied here. A penalty-adjusted cluster-robust covariance estimator yields tests with near-nominal size for finite-dimensional parameters, and confidence bands provide accurate coverage for centered unknown functions.
\end{abstract}


\jel{C14, C15, C23, C87}

\newpage

\section{Introduction}\label{sec:introduction}

Empirical work in economics frequently uses panel data to account for unobserved unit-level heterogeneity through fixed effects. At the same time, many applied questions call for flexible functional forms. For example, \citet{borri_martins_filho_kalatzis_2022} estimate a nonlinear investment curve, while \citet{baglan_yoldas_2014} estimate an inflation--growth relationship semiparametrically. Such empirical applications remain relatively rare, however, despite a well-developed theoretical literature on semiparametric and nonparametric panel data models.

A central practical obstacle is the need to handle fixed effects while also choosing basis dimension, series length, bandwidths, or smoothing penalties. Nonparametric methods that are popular in cross-sectional settings, such as kernel or nearest-neighbor estimators, do not naturally accommodate additive fixed effects. Series-based approaches can incorporate fixed effects more directly, but they require the researcher to specify basis functions, choose interaction terms, and select the number of series terms.

This paper provides a practical guide to estimating semiparametric and nonparametric fixed effects panel data models using penalized splines. The approach connects three literatures. The first is the econometric literature on series estimation for semiparametric panel data models with fixed effects \citep{baltagi_li_2002,an_et_al_2016}. The second is the broader econometric literature on sieve and penalized sieve estimation and inference for nonparametric and semiparametric models \citep{ai_chen_2003,chen_2007,chen_pouzo_2009,chen_pouzo_2012,chen_pouzo_2015}. The third is the statistical literature on generalized additive models, smoothing parameter selection, and penalized regression splines \citep{wood_2003,wood_2011,wood_2017_gam}.

Conceptually, the proposed approach is grounded in the econometric theory of series, sieve, and penalized sieve estimation. Computationally, it uses the fast and stable implementations developed in the generalized additive model literature, making flexible estimation practical even in panel settings with high-dimensional fixed effects.

The paper gives concrete implementation guidance in R using the \code{mgcv} package \citep{wood_mgcv_package}. I first describe the class of fully nonparametric, partially linear, and varying-coefficient panel models covered by the approach. I then show how the resulting penalized-spline estimators can be implemented using unit indicators or first differences, how a penalized unit-effect specification can be used as a regularized alternative, and how to construct penalty-adjusted cluster-robust standard errors and confidence bands from the fitted \code{bam} object. Monte Carlo simulations evaluate estimation accuracy for the unknown function, empirical size and power for finite-dimensional parameters, and coverage of confidence bands for the nonparametric component.

\section{Models}\label{sec:models}

The proposed approach can handle a variety of fixed effects panel data models. One example is the fully nonparametric model
\begin{align}\label{eq:np_fe_model}
Y_{it} = g(X_{it}) + \mu_i + \varepsilon_{it}, \qquad
E[ \varepsilon_{it} \mid X_{i1}, \ldots, X_{iT}, \mu_i ] = 0,
\end{align}
where \(X_{it}\) may be scalar or vector-valued, \(g(\cdot)\) is unknown, and \(\mu_i\) is a unit fixed effect.

Another example is the partially linear model
\begin{align}\label{eq:pl_fe_model}
Y_{it} = X_{1,it}'\beta + g(X_{2,it}) + \mu_i + \varepsilon_{it}, \qquad
E[ \varepsilon_{it} \mid X_{i1}, \ldots, X_{iT}, \mu_i ] = 0,
\end{align}
where \(X_{it} = (X_{1,it}', X_{2,it}')'\), \(\beta\) is an unknown finite-dimensional parameter, and \(g(\cdot)\) is an unknown function.

A third example is a varying-coefficient model,
\begin{align}\label{eq:vc_fe_model}
Y_{it} = \beta_0(X_{2,it}) + X_{1,it}'\beta_1(X_{2,it}) + \mu_i + \varepsilon_{it}, \qquad
E[ \varepsilon_{it} \mid X_{i1}, \ldots, X_{iT}, \mu_i ] = 0,
\end{align}
where \(\beta_0(\cdot)\) is an unknown intercept function and \(\beta_1(\cdot)\) is an unknown smooth coefficient function, or vector of coefficient functions when \(X_{1,it}\) is vector-valued.

In these additive fixed-effects models, the level of the unknown smooth is not separately identified from the intercept and unit effects. The identified object is a normalized version of the function. For population statements one can write \(g^c(x)=g(x)-E[g(X_{2,it})]\), with the obvious modification in the fully nonparametric model. The tables below use the operational grid analog
\[
g_G^c(x_j)=g(x_j)-\frac{1}{J}\sum_{\ell=1}^J g(x_\ell),
\qquad j=1,\ldots,J,
\]
on the evaluation grid \(G=\{x_1,\ldots,x_J\}\). For RMSE calculations the same normalization is applied on the independent test set. For first-differenced estimators, the data identify smooth contrasts \(g(X_{2,it})-g(X_{2,i,t-1})\); reported level curves are reconstructed in the fitted basis and then normalized using the same grid-centering convention.

\section{Estimation: Penalized Objective}\label{sec:estimation}

This section describes the penalized objective that underlies the estimators used in this paper and the role of fast restricted maximum likelihood (fREML) in selecting the penalty weights. A Gaussian identity-link \code{gam} or \code{bam} fit can be viewed as a penalized least squares problem with the amount of penalization chosen automatically. After each smooth term has been represented by a finite set of basis functions, the model is linear in its unknown coefficients. Let $R$ denote the resulting \code{mgcv} model matrix, including parametric regressors, spline basis columns, and any factor indicators, and let $\theta$ collect all coefficients. For a fixed vector of smoothing parameters $\lambda=(\lambda_1,\ldots,\lambda_J)'$, \code{mgcv} estimates
\begin{align}\label{eq:mgcv_penalized_objective}
\hat{\theta}_{\lambda}
&=
\argmin_{\theta}
Q_{\lambda}(\theta), \\
Q_{\lambda}(\theta)
&=
\sum_{i=1}^n\sum_{t=1}^T
\left(Y_{it}-R_{it}'\theta\right)^2
+
\sum_{j=1}^J \lambda_j \theta' S_j \theta \notag \\
&=
\|Y-R\theta\|^2+\theta'S_{\lambda}\theta,
\qquad
S_{\lambda}=\sum_{j=1}^J \lambda_j S_j . \notag
\end{align}
The matrices $S_j$ are known positive semidefinite penalty matrices constructed from the chosen smooth bases; each $S_j$ has nonzero entries only for the coefficients belonging to the corresponding smooth term. This is directly analogous to the familiar ridge-type objective
\[
\min_{\beta_0,\ldots,\beta_k}
\sum_i \left(Y_i - \beta_0 - \sum_{j=1}^{k} X_{ij}\beta_j\right)^2
+ \lambda \sum_{j=1}^{k}\beta_j^2.
\]
Here the intercept is left unpenalized, while larger values of \(\lambda\) shrink the slope coefficients toward zero. In the spline case, however, the row \(R_i\) contains the usual regressors, spline basis functions, fixed-effect indicators, and any other columns generated by the model formula, rather than only the raw covariates \(X_i\).

For common cubic and thin-plate spline bases, the quadratic form $\theta'S_j\theta$ is a finite-dimensional measure of wiggliness, such as an integrated squared derivative. Thus \(\lambda_j\) controls how costly wiggliness is for the \(j\)th smooth. A large $\lambda_j$ pushes the smooth toward the null space of its penalty, such as a low-order polynomial component; with \code{select = TRUE}, \code{mgcv} adds an additional shrinkage penalty on this null space, so that a smooth can be penalized all the way toward zero. Ordinary parametric regressors and factor fixed effects are not penalized unless an explicit parametric penalty is supplied. By contrast, a term such as \code{s(unit, bs = "re")} uses a ridge-type penalty on the unit effects, corresponding to a Gaussian random-effect representation.

The label \code{bs = "re"} refers to a computational random-effect representation of a ridge penalty on the unit indicators, not to the classical random-effects assumption that the unit effect is uncorrelated with the regressors. Conditional on the smoothing parameter, the estimator is penalized least squares with a quadratic penalty on the unit effects. Because this specification shrinks the unit effects toward zero, however, the resulting estimator differs numerically from including unpenalized factor fixed effects. When the unit effects are correlated with the regressors and \(T\) is fixed, this shrinkage can transmit part of that correlation into the estimated smooth or finite-dimensional coefficient. I therefore treat \code{s(unit, bs = "re")} as a predictive or regularized alternative, not as a substitute for the fixed-effects estimand delivered by unpenalized unit indicators or first differencing.

It is important to distinguish the coefficient objective in equation~\eqref{eq:mgcv_penalized_objective} from the role of \code{method = "REML"} or \code{method = "fREML"}. Conditional on \(\lambda\), the coefficient estimates solve the penalized least squares problem above; REML or fREML is then used to choose \(\lambda\) and the scale parameter. In the Gaussian case, this can be understood by treating the penalized spline coefficients as Gaussian random effects with precision proportional to the smoothing parameters. For a trial value of \(\lambda\), \code{mgcv} solves the penalized coefficient problem and evaluates a restricted marginal likelihood criterion, with the scale parameter profiled or updated. The outer smoothing-parameter step then searches over \(\lambda\). Schematically, up to constants and rank adjustments, the negative REML criterion has the form
\[
\mathcal{V}_{\mathrm{REML}}(\lambda,\sigma^2)
=
\frac{1}{2\sigma^2}Q_{\lambda}(\hat{\theta}_{\lambda})
+\frac{1}{2}\log |R'R+S_{\lambda}|
-\frac{1}{2}\log |S_{\lambda}|_{+}
+\frac{N-d_0}{2}\log \sigma^2,
\]
where $N=nT$, $|\cdot|_+$ denotes the product of positive eigenvalues, and $d_0$ is the dimension of the unpenalized component after imposing the relevant identifiability constraints. The data-fit contribution enters through \(Q_{\lambda}(\hat{\theta}_{\lambda})\) and the scale term. The determinant terms adjust the marginal likelihood for the design, penalty, and dimension of the penalized and unpenalized coefficient spaces, thereby discouraging overly flexible smooths when the smoothing parameters are chosen. The \code{bam} option \code{fREML} is a computationally efficient REML-type implementation for large data sets \citep{wood_2011,wood_goude_shaw_2015,wood_li_shaddick_augustin_2017}. Thus, for the estimators used below, the coefficient objective is ``RSS plus a quadratic wiggliness penalty,'' while REML/fREML selects the penalty weights rather than requiring the researcher to choose them by hand.

While REML/fREML is the default choice in \code{mgcv::bam}, one could use a more familiar generalized cross-validation (GCV) criterion. REML/fREML does not require a fully parametric likelihood model for the substantive econometric object of interest; instead, it is used as a data-driven rule for selecting the smoothing parameters. Other smoothing selection criteria, including GCV/Cp-type criteria, are also available in \code{mgcv}. I use fREML in the simulations because REML-type criteria are often more stable in finite samples and less prone to undersmoothing than GCV-type criteria \citep{reiss_ogden_2009}, but the same implementation framework can be used with GCV if desired.

\section{Inference}\label{sec:inference}

Standard \code{mgcv} returns a Bayesian or frequentist covariance matrix that assumes independent, homoskedastic observations. For panel data with within-unit dependence, a cluster-robust analogue is needed. This section develops a penalty-adjusted cluster-robust covariance estimator from the fitted \code{bam} object.

Let $R$ denote the fitted \code{mgcv} linear predictor matrix, $S_{\lambda}$ the penalty matrix, and $\hat{u}$ the residual vector. For Gaussian identity-link models without observation weights, the \code{mgcv} working-weight matrix is the identity matrix, so the penalized Hessian matrix is
\begin{align}\label{eq:B_lambda}
B_{\lambda}=R'R+S_{\lambda}.
\end{align}
A penalty-adjusted clustered sandwich then has the form
\begin{align}\label{eq:penalty_cluster_sandwich}
\widehat{V}
=
B_{\lambda}^{-1}
\hat{\Omega}
B_{\lambda}^{-1}, \quad 
\hat{\Omega} = 
\sum_{i=1}^n
\hat{s}_i \hat{s}_i'
, \quad 
\hat{s}_i = \sum_{t=1}^T R_{it}'\hat{u}_{it},
\end{align}
where $R_{it}$ is the row of $R$ for observation $(i,t)$. $\hat{\Omega}$ is the unit-cluster robust covariance component: it first sums the score contributions $R_{it}'\hat{u}_{it}$ over time within each unit and then forms outer products of the unit-level sums. This has the same unit-level outer-product form as the robust covariance estimator for within-group estimators in \citet{arellano_1987}, and therefore allows arbitrary heteroskedasticity and serial dependence within units under independence across units. In the present setting, this expression can be viewed as a penalized-spline analog of the score covariance term in series-based semiparametric panel estimators. Setting $S_{\lambda}=0$ gives the corresponding series-style cluster sandwich that ignores the penalty.

\subsection{Inference on Finite-Dimensional Components}

For a scalar finite-dimensional parameter \(\beta\), or for one component of a vector \(\beta\), let \(\hat{\sigma}_{\beta}^2\) denote the corresponding diagonal element of \(\widehat{V}\). Under the Gaussian approximation,
\[
\frac{\hat{\beta}-\beta}{\hat{\sigma}_{\beta}}
\approx N(0,1).
\]
Thus an approximate \(1-\alpha\) confidence interval for \(\beta\) is
\[
\left[
\hat{\beta}-z_{1-\alpha/2}\hat{\sigma}_{\beta},\,
\hat{\beta}+z_{1-\alpha/2}\hat{\sigma}_{\beta}
\right].
\]

\subsection{Inference on the Nonparametric Component}

For inference on the nonparametric component, let \(G=\{x_1,\ldots,x_J\}\) be the evaluation grid. Let \(R_G\) denote the grid \code{lpmatrix}, and let \(R_G^{(g)}\) be the same matrix after all columns not belonging to the smooth of interest have been set to zero. Because the level of \(g\) is not separately identified from the intercept and unit effects, the target is the grid-centered smooth. Let
\[
C_G=I_J-\mathbbm{1}\mathbbm{1}'/J
\]
be the grid-centering matrix, where \(\mathbbm{1}\) is the \(J\)-vector of ones. Premultiplication by \(C_G\) subtracts the average over the \(J\) grid points. The matrix that maps fitted coefficients into centered smooth values is
\[
L_G=C_G R_G^{(g)}.
\]
Thus
\[
\hat{g}_G^c=L_G\hat{\theta},
\qquad
\hat{g}_G^c(x_j)
=
\hat{g}(x_j)-\frac{1}{J}\sum_{\ell=1}^J \hat{g}(x_\ell),
\]
and the estimated covariance matrix for the centered grid vector is
\[
\widehat{\Sigma}_G=L_G\widehat{V}L_G'.
\]
Pointwise and uniform bands use the same estimated grid covariance matrix but answer different questions. A pointwise \(1-\alpha\) interval at grid point \(x_j\) is
\[
\hat{g}_G^c(x_j)\pm z_{1-\alpha/2}\hat{\sigma}_j,
\qquad
\hat{\sigma}_j^2=(\widehat{\Sigma}_G)_{jj}.
\]
This interval is calibrated separately at each \(x_j\). For each individual point,
\[
P\left(
g_G^c(x_j)
\in
\left[
\hat{g}_G^c(x_j)-z_{1-\alpha/2}\hat{\sigma}_j,\,
\hat{g}_G^c(x_j)+z_{1-\alpha/2}\hat{\sigma}_j
\right]
\right)
\approx 1-\alpha.
\]
This pointwise statement does not imply that all intervals cover the curve simultaneously. The distinction between pointwise component intervals and whole-function coverage is standard in spline and GAM inference; see, for example, \citet{marra_wood_2012} on coverage of GAM component intervals.

A uniform band instead controls the probability that the entire reported grid curve is covered simultaneously:
\[
P\left(
g_G^c(x_j)\in
\left[\hat{g}_G^c(x_j)-c_{1-\alpha}\hat{\sigma}_j,\,
\hat{g}_G^c(x_j)+c_{1-\alpha}\hat{\sigma}_j\right]
\text{ for all } j=1,\ldots,J
\right)\approx 1-\alpha.
\]
The critical value \(c_{1-\alpha}\) is the \(1-\alpha\) quantile of the maximum absolute standardized error over the grid. Each standardized component is marginally standard normal, but the maximum depends on their joint correlation structure. For a spline fit, nearby grid points are typically strongly correlated, so the maximum is neither the maximum of independent standard normals nor a single standard normal. Except in special cases, there is no simple scalar closed-form critical value; simultaneous bands for penalized splines are therefore commonly constructed from the approximating Gaussian distribution \citep{krivobokova_kneib_claeskens_2010}. In the simulations I approximate this critical value from an auxiliary Gaussian vector \(Z_G\sim N(0,\widehat{\Sigma}_G)\), where \(Z_G\) is simulated rather than observed in the data:
\[
c_{1-\alpha}
=
\operatorname{quantile}_{1-\alpha}
\left(
\max_{1\leq j\leq J}
\left|Z_{G,j}/\hat{\sigma}_j\right|
\right).
\]
This is a simultaneous band on the reported grid. Coverage is evaluated over the 50-point grid used in the simulation.

\section{Implementation Guide}\label{sec:implementation}

This section translates the preceding estimation and inference formulas into \code{mgcv} code. Suppose that one wants to estimate the model from equation~\ref{eq:pl_fe_model}, with scalar $X_1$ and $X_2$. There are several ways to implement the estimator, and the same post-estimation covariance calculations can then be applied to the fitted object.

\subsection{Estimation}

\subsubsection{Fixed Effects}

The main implementation uses factor fixed effects:
\begin{lstlisting}
df$unit <- factor(df$unit)
bam(y ~ x1 + s(x2, k = 20) + unit,
    data = df,
    method = "fREML",
    select = TRUE,
    discrete = TRUE)
\end{lstlisting}

\subsubsection{Penalized Unit Effects}

An alternative regularized specification uses a penalized unit effect:
\begin{lstlisting}
bam(y ~ x1 + s(x2, k = 20) + s(unit, bs = "re"),
    data = df,
    method = "fREML",
    select = TRUE,
    discrete = TRUE)
\end{lstlisting}
This specification is often useful for prediction or as a shrinkage comparison, but it changes the estimator when \(\mu_i\) is correlated with the regressors. The fixed-effects interpretation in the usual panel-data sense comes from the factor-indicator or first-differenced specifications.

\subsubsection{First Differences}

Fixed effects can also be removed by first differencing. A constrained smooth of the form
\[
g(X_{it}) - g(X_{i,t-1})
\]
can be represented in \code{mgcv} using linear functional smooths. The implementation constructs the matrix of current and lagged covariates and the corresponding $+1/-1$ weights.

The core first-differenced \code{bam} call used in the simulations is:
\begin{lstlisting}
X_pair <- cbind(df_fd$x2_now, df_fd$x2_lag)
L_pair <- cbind(rep(1, nrow(df_fd)), rep(-1, nrow(df_fd)))

bam(dy ~ dx1 + s(X_pair, by = L_pair, k = 20),
    data = list(
      dy = df_fd$dy,
      dx1 = df_fd$dx1,
      X_pair = X_pair,
      L_pair = L_pair
    ),
    method = "fREML",
    select = TRUE,
    discrete = TRUE)
\end{lstlisting}

Here \code{dy} and \code{dx1} are first differences, while \code{X\_pair} stores $(X_{2,it},X_{2,i,t-1})$ and \code{L\_pair} stores the corresponding $(1,-1)$ weights. The term \code{s(X\_pair, by = L\_pair)} therefore estimates a single smooth $g(\cdot)$ evaluated at both arguments and constrained to enter as $g(X_{2,it})-g(X_{2,i,t-1})$.

\subsubsection{Additional Options}

By default, \code{s()} smooths in \code{mgcv} use thin plate regression spline bases, specified by \code{bs = "tp"} \citep{wood_2003}. Researchers can also choose other spline bases, such as cubic regression splines with \code{bs = "cr"} or P-splines with \code{bs = "ps"}.

The same syntax extends naturally to varying-coefficient models. For example, a model in which the coefficient on \(X_{1,it}\) varies smoothly with \(X_{2,it}\),
\[
Y_{it} = \beta_0(X_{2,it}) + X_{1,it} \beta_1(X_{2,it}) + \mu_i + \varepsilon_{it},
\]
can be estimated using a by-variable smooth:
\begin{lstlisting}
bam(y ~ s(x2, k = 20) + s(x2, by = x1, k = 20) + unit,
    data = df,
    method = "fREML",
    select = TRUE,
    discrete = TRUE)
\end{lstlisting}
The simulations below focus on one-dimensional smooths; the varying-coefficient and tensor-product examples are included to show how the same fixed-effects syntax extends to richer specifications.

For a fully nonparametric bivariate component \(g(X_{1,it},X_{2,it})\), one can use a tensor-product smooth:
\begin{lstlisting}
bam(y ~ te(x1, x2, k = c(10, 10)) + unit,
    data = df,
    method = "fREML",
    select = TRUE,
    discrete = TRUE)
\end{lstlisting}
If the goal is to separate main effects from interactions, \code{ti()} provides the corresponding tensor-product interaction:
\begin{lstlisting}
bam(y ~ s(x1, k = 10) + s(x2, k = 10) + ti(x1, x2, k = c(10, 10)) + unit,
    data = df,
    method = "fREML",
    select = TRUE,
    discrete = TRUE)
\end{lstlisting}

The basis dimension \(k\) sets the maximum flexibility of the smooth and is chosen by the researcher; conditional on this basis, \code{mgcv} chooses the smoothing parameter by fREML. If \(k\) is omitted, \code{mgcv} uses default basis dimensions, but these defaults should be checked in applications.

For \code{bam}, the default smoothing parameter selection criterion is fREML, corresponding to \code{method = "fREML"}. Researchers who wish to use the GCV/Cp criterion instead can specify \code{method = "GCV.Cp"}. The default choices are best viewed as convenient and computationally efficient options, rather than binding or unusual constraints imposed on the estimator.

\subsection{Inference}

After fitting any of the fixed-effects, penalized-unit-effects, or first-differenced specifications, standard errors are computed from the fitted bam object rather than from the default mgcv covariance matrix. The fitted linear predictor matrix is obtained with
\begin{lstlisting}
R <- predict(fit, type = "lpmatrix")
u <- residuals(fit, type = "response")
\end{lstlisting}
and the unit-level score sums are formed by summing the row-wise products of the design matrix and residuals within unit:
\begin{lstlisting}
score_g <- rowsum(R * as.numeric(u),
                  group = as.factor(unit_id),
                  reorder = FALSE)
Omega_hat <- crossprod(score_g)
\end{lstlisting}
This command produces $\hat{\Omega}$ from equation~\eqref{eq:penalty_cluster_sandwich}.

For the penalty-adjusted covariance estimator, one needs to obtain $B_{\lambda}^{-1}$ from equation~\eqref{eq:B_lambda}. In the Gaussian identity-link case this is obtained from the Bayesian covariance matrix returned by \code{mgcv}, rescaled by the estimated residual variance:
\begin{lstlisting}
B_lambda_inv <- vcov(fit, freq = FALSE) / summary(fit)$scale
V_penalty <- B_lambda_inv %*% Omega_hat %*% B_lambda_inv
\end{lstlisting}
The simulations also apply the usual finite-sample cluster correction for a one-way cluster-robust covariance matrix \citep{cameron_miller_2015}. In this panel setting, each unit is one cluster, so the number of clusters is \(n\). Let \(N\) denote the total number of observations and let \(d\) denote the substantive degrees of freedom of the fitted model. The correction used below is
\[
\frac{n}{n-1}\frac{N-1}{N-d}.
\]
For a balanced panel, \(N=nT\); for an unbalanced panel, \(N\) is simply the total number of observed unit-time cells. For the fixed-effects specification, \(d\) excludes nuisance unit indicators, matching the within-estimator convention rather than inflating the adjustment by counting every unit effect. In code, the correction is:
\begin{lstlisting}
is_unit_effect_coef <- function(coef_names) {
  startsWith(coef_names, "unit") |
    startsWith(coef_names, "s(unit).")
}

substantive_edf <- function(fit) {
  coef_names <- names(fit$edf)
  keep <- !is_unit_effect_coef(coef_names)
  sum(fit$edf[keep])
}

cluster_correction <- function(fit, score_g, R) {
  n_clusters <- nrow(score_g)
  nobs <- nrow(R)
  df_used <- substantive_edf(fit)

  (n_clusters / (n_clusters - 1)) *
    ((nobs - 1) / max(nobs - df_used, 1))
}

corr <- cluster_correction(fit, score_g, R)
V_penalty <- corr * V_penalty
\end{lstlisting}

For pointwise and uniform bands for $g(\cdot)$, the fitted smooth is evaluated on a grid by calling \code{predict(..., type = "lpmatrix")} on the grid data. As in Section~\ref{sec:inference}, the full prediction matrix should not be used unchanged, because it includes the intercept, linear regressors, and unit effects. In code, the construction of \(L_G=C_G R_G^{(g)}\) corresponds to selecting the columns for \code{s(x2)}, zeroing the remaining columns, and centering the selected grid rows:
\begin{lstlisting}
grid_df <- data.frame(
  x1 = 0,
  x2 = x_grid,
  unit = factor(reference_unit, levels = levels(df$unit))
)
Rg <- predict(fit, newdata = grid_df, type = "lpmatrix")

smooth_id <- which(vapply(fit$smooth, `[[`, "", "label") == "s(x2)")
sm <- fit$smooth[[smooth_id[1]]]
smooth_cols <- sm$first.para:sm$last.para

L <- Rg
L[, -smooth_cols] <- 0
L <- sweep(L, 2, colMeans(L), "-")

g_hat <- as.numeric(L %*% coef(fit))
C_grid <- L %*% V_penalty %*% t(L)
\end{lstlisting}
Pointwise intervals only require the diagonal of \code{C\_grid}:
\begin{lstlisting}
se_grid <- sqrt(pmax(diag(C_grid), 0))
z_crit <- qnorm(0.975)

point_lower <- g_hat - z_crit * se_grid
point_upper <- g_hat + z_crit * se_grid
\end{lstlisting}
These intervals are appropriate for statements about a fixed grid point, such as the value of the centered smooth at \(x=0.5\). A plotted collection of pointwise intervals is visually useful, but it is not a simultaneous 95\% band for the curve.

For a uniform band, first convert the covariance matrix into the correlation matrix of the standardized grid errors, simulate draws from the corresponding Gaussian approximation, and take the 95th percentile of the maximum absolute draw. The simulations below use \(499\) Gaussian draws for this step:
\begin{lstlisting}
valid <- is.finite(se_grid) & se_grid > 1e-10
Corr <- C_grid[valid, valid, drop = FALSE] /
  outer(se_grid[valid], se_grid[valid])
Corr <- (Corr + t(Corr)) / 2

eig <- eigen(Corr, symmetric = TRUE)
vals <- pmax(eig$values, 0)
A <- eig$vectors %*% diag(sqrt(vals), nrow = length(vals))

n_draws <- 499
sim <- matrix(rnorm(n_draws * length(vals)), nrow = n_draws) %*% t(A)
sup_crit <- as.numeric(
  quantile(apply(abs(sim), 1, max), 0.95, names = FALSE)
)

uniform_lower <- g_hat - sup_crit * se_grid
uniform_upper <- g_hat + sup_crit * se_grid
\end{lstlisting}
The only difference between the pointwise and uniform bands is the critical value: \(1.96\) for each separate point versus the simulated sup-norm critical value for the whole grid. The number of Gaussian draws controls only the simulation error in this critical value; using more draws, such as \(999\) or \(1{,}999\), gives a slightly more stable cutoff at additional computational cost. In the first-differenced implementation, the grid data evaluate the linear functional smooth with \code{X\_pair = x\_grid} and \code{L\_pair = 1}; the same zeroing, centering, and pointwise-or-uniform critical-value steps then report the normalized level curve \(g_G^c\).

\section{Simulation Design}\label{sec:simulations}

The simulation section studies both estimation accuracy and inference. For estimation of $g(\cdot)$, the benchmark loss is out-of-sample root mean squared error on an independently generated test set. For inference on $\beta$, the benchmark outcomes are empirical size under the null and empirical power under the alternative. For inference on $g(\cdot)$, the benchmark outcome is coverage of confidence bands.

For inference on $g(\cdot)$, I primarily use a penalty-adjusted cluster sandwich estimator. A series-style cluster sandwich estimator that ignores the penalty is too conservative, so its results are reported in the appendix. One could consider a wild cluster bootstrap that refits the \code{bam} model in each bootstrap draw, but it is left for future work.

The simulation designs deliberately correlate the unit effects with the regressors. For each unit, draw a latent index \(\alpha_i\sim \mathrm{Beta}(2.2,2.2)\) and set \(\mu_i=1.1(\alpha_i-\bar{\alpha})+\nu_i\), where \(\nu_i\sim N(0,1.25^2)\) is independent noise. In the one-regressor designs, \(X_{it}\) is a noisy draw around \(\alpha_i\), clipped to lie in $[0, 1]$. In the partially linear designs, \(X_{2,it}\) is generated the same way and \(X_{1,it}\) is correlated with both \(X_{2,it}\) and \(\alpha_i\). Thus the unobserved unit effect is not independent of the covariates.

The tables use a heteroskedastic AR(1) error design as the primary specification. Conditional variances are proportional to \(1+2\exp(0.75 Z_{it})\) and normalized to keep the average error variance fixed across designs. Here \(Z_{it}\) denotes the covariate index used in the variance function. In the estimation-accuracy designs and nonparametric coverage simulations, there is a single nonparametric regressor, so \(Z_{it}=X_{it}\). The partially linear outcome adds \(X_{1,it}\beta\), and for that design \(Z_{it}=X_{1,it}+X_{2,it}\). Within units, errors have AR(1) correlation with \(\rho=0.5\). All reported tables use \(M=1{,}000\) Monte Carlo draws.

\subsection{Estimation Accuracy}

I consider three data generating processes (DGPs):
\begin{align}
g_1(x) &= 0.5 + 1.4x,\\
g_2(x) &= 0.5 + 1.4x + 0.75(x-0.5)^2,\\
g_3(x) &= \sin(2\pi x),
\end{align}
and the outcome is then generated as
\[
Y_{it} = g_j(X_{it}) + \mu_i + \varepsilon_{it}.
\]

Table~\ref{tab:estimation_rmse} reports estimation accuracy for the three sample-size designs. Each entry reports $100$ times the root mean squared error for the centered function on an independently generated test set with $1{,}000$ test points per draw. The fixed spline estimators use $K=7$ basis terms. This fixed-\(K\) benchmark can perform well when its number of series terms ``matches'' the smoothness of the unknown function, but is unable to adapt to the smoothness automatically. In the sine design, the fixed-\(K\) spline performs especially well, but it overfits in the other two designs and is dominated by the \code{bam} estimators.

\begin{table}[H]
\centering
\caption{Out-of-sample RMSE for estimating $g$}
\label{tab:estimation_rmse}
\begin{tabular}{llrrr}
\toprule
DGP & Method & \multicolumn{3}{c}{Mean RMSE ($\times 100$)} \\
\cmidrule(lr){3-5}
& & $n=200$, $T=4$ & $n=200$, $T=8$ & $n=500$, $T=4$ \\
\midrule
linear & linear FE & 1.035 & 0.772 & 0.641 \\
linear & linear FD & 0.982 & 0.680 & 0.632 \\
linear & spline FE, fixed $K$ & 2.378 & 1.704 & 1.489 \\
linear & spline FD, fixed $K$ & 2.247 & 1.504 & 1.433 \\
linear & \code{bam} FD & 1.182 & 0.807 & 0.757 \\
linear & \code{bam} FE & 1.186 & 0.882 & 0.744 \\
linear & \code{bam} RE & 1.213 & 0.884 & 0.795 \\
\midrule
quadratic & linear FE & 5.985 & 5.906 & 5.892 \\
quadratic & linear FD & 5.986 & 5.895 & 5.888 \\
quadratic & spline FE, fixed $K$ & 2.369 & 1.713 & 1.499 \\
quadratic & spline FD, fixed $K$ & 2.255 & 1.511 & 1.425 \\
quadratic & \code{bam} FD & 1.713 & 1.204 & 1.136 \\
quadratic & \code{bam} FE & 1.767 & 1.320 & 1.169 \\
quadratic & \code{bam} RE & 1.781 & 1.325 & 1.211 \\
\midrule
$\sin(2\pi x)$ & linear FE & 48.076 & 47.996 & 47.951 \\
$\sin(2\pi x)$ & linear FD & 48.122 & 48.015 & 47.968 \\
$\sin(2\pi x)$ & spline FE, fixed $K$ & 2.366 & 1.723 & 1.510 \\
$\sin(2\pi x)$ & spline FD, fixed $K$ & 2.270 & 1.541 & 1.450 \\
$\sin(2\pi x)$ & \code{bam} FD & 2.625 & 1.904 & 1.807 \\
$\sin(2\pi x)$ & \code{bam} FE & 2.678 & 2.052 & 1.834 \\
$\sin(2\pi x)$ & \code{bam} RE & 2.692 & 2.056 & 1.862 \\
\bottomrule
\end{tabular}
\begin{minipage}{0.92\textwidth}
\footnotesize \tablenote\ Entries report $100$ times the root mean squared error for the centered function on an independent test set. All designs use heteroskedastic AR(1) errors with \(\rho=0.5\).
\end{minipage}
\end{table}

\subsection{Inference}

The semiparametric inference simulations use the partially linear model
\[
Y_{it} = X_{1,it}\beta + g(X_{2,it}) + \mu_i + \varepsilon_{it}.
\]
The first exercise reports empirical rejection probabilities for tests of $H_0:\beta=1$, both under the null and under a fixed alternative.

Table~\ref{tab:beta_size} reports size results for tests of $H_0:\beta=1$ in the partially linear model. The design uses $g(x)=\sin(2\pi x)$ and the heteroskedastic AR(1) error process described above.

\begin{table}[H]
\centering
\caption{Size results for tests of $H_0:\beta=1$}
\label{tab:beta_size}
\begin{tabular}{lrrrrr}
\toprule
Method & Classic & Penalty & Mean $\hat{\beta}$ & SD $\hat{\beta}$ & Mean SE \\
\midrule
\multicolumn{6}{l}{\textit{Panel A: $n=200$, $T=4$}} \\
\midrule
\code{bam} FD & 0.060 & 0.063 & 1.0003 & 0.0373 & 0.0372 \\
\code{bam} FE & 0.051 & 0.054 & 0.9999 & 0.0380 & 0.0385 \\
\code{bam} RE & 0.046 & 0.051 & 1.0037 & 0.0378 & 0.0381 \\
\midrule
\multicolumn{6}{l}{\textit{Panel B: $n=200$, $T=8$}} \\
\midrule
\code{bam} FD & 0.056 & 0.058 & 0.9994 & 0.0258 & 0.0250 \\
\code{bam} FE & 0.059 & 0.062 & 0.9992 & 0.0287 & 0.0278 \\
\code{bam} RE & 0.052 & 0.056 & 1.0012 & 0.0286 & 0.0277 \\
\midrule
\multicolumn{6}{l}{\textit{Panel C: $n=500$, $T=4$}} \\
\midrule
\code{bam} FD & 0.056 & 0.059 & 0.9999 & 0.0240 & 0.0236 \\
\code{bam} FE & 0.051 & 0.051 & 1.0006 & 0.0244 & 0.0244 \\
\code{bam} RE & 0.051 & 0.054 & 1.0043 & 0.0243 & 0.0242 \\
\bottomrule
\end{tabular}
\begin{minipage}{0.92\textwidth}
\footnotesize \tablenote\ The columns ``Classic'' and ``Penalty'' report empirical rejection frequencies using the cluster sandwich that ignores the penalty and the penalty-adjusted cluster sandwich, respectively. ``Mean SE'' reports the mean penalty-adjusted standard error.
\end{minipage}
\end{table}

Table~\ref{tab:beta_power} reports power results for the same test when the true value is \(\beta=1.075\).

\begin{table}[H]
\centering
\caption{Power results for tests of $H_0:\beta=1$}
\label{tab:beta_power}
\begin{tabular}{lrrrrr}
\toprule
Method & Classic & Penalty & Mean $\hat{\beta}$ & SD $\hat{\beta}$ & Mean SE \\
\midrule
\multicolumn{6}{l}{\textit{Panel A: $n=200$, $T=4$}} \\
\midrule
\code{bam} FD & 0.515 & 0.525 & 1.0753 & 0.0373 & 0.0372 \\
\code{bam} FE & 0.478 & 0.485 & 1.0749 & 0.0380 & 0.0385 \\
\code{bam} RE & 0.525 & 0.545 & 1.0787 & 0.0378 & 0.0381 \\
\midrule
\multicolumn{6}{l}{\textit{Panel B: $n=200$, $T=8$}} \\
\midrule
\code{bam} FD & 0.829 & 0.832 & 1.0744 & 0.0258 & 0.0250 \\
\code{bam} FE & 0.756 & 0.758 & 1.0742 & 0.0287 & 0.0278 \\
\code{bam} RE & 0.772 & 0.779 & 1.0762 & 0.0286 & 0.0277 \\
\midrule
\multicolumn{6}{l}{\textit{Panel C: $n=500$, $T=4$}} \\
\midrule
\code{bam} FD & 0.872 & 0.873 & 1.0749 & 0.0240 & 0.0236 \\
\code{bam} FE & 0.870 & 0.872 & 1.0756 & 0.0244 & 0.0244 \\
\code{bam} RE & 0.909 & 0.919 & 1.0793 & 0.0243 & 0.0242 \\
\bottomrule
\end{tabular}
\begin{minipage}{0.92\textwidth}
\footnotesize \tablenote\ The columns ``Classic'' and ``Penalty'' report empirical rejection frequencies using the cluster sandwich that ignores the penalty and the penalty-adjusted cluster sandwich, respectively. The null is \(H_0:\beta=1\), while the true value is \(\beta=1.075\). ``Mean SE'' reports the mean penalty-adjusted standard error.
\end{minipage}
\end{table}

Tables~\ref{tab:g_coverage_pl_penalty} and~\ref{tab:g_coverage_np_penalty} report penalty-adjusted inference for the unknown function $g$. Table~\ref{tab:g_coverage_pl_penalty} continues with the partially linear model used for the \(\beta\) results, now focusing on the smooth component. Table~\ref{tab:g_coverage_np_penalty} reports the corresponding results for the fully nonparametric model \(Y_{it}=g(X_{it})+\mu_i+\varepsilon_{it}\). In both cases, the target is the centered function evaluated on a grid of 50 points, with $g(x)=\sin(2\pi x)$. The ``Avg. pointwise'' column averages marginal coverage over the grid points, while ``Min. pointwise'' reports the worst grid-point coverage. The ``Uniform'' column reports the fraction of Monte Carlo draws in which the entire grid curve is covered simultaneously by the sup-$t$ band, using \(499\) Gaussian draws to approximate the critical value in each fitted model. The corresponding series-style bands that ignore the penalty are reported in Appendix Table~\ref{tab:g_coverage_classic_appendix}.

\begin{table}[H]
\centering
\caption{Penalty-adjusted coverage results for centered $g$: partially linear model}
\label{tab:g_coverage_pl_penalty}
\begin{tabular}{lrrrr}
\toprule
Method & Avg. pointwise & Min. pointwise & Uniform & Avg. width \\
\midrule
\multicolumn{5}{l}{\textit{Panel A: $n=200$, $T=4$}} \\
\midrule
\code{bam} FD & 0.943 & 0.927 & 0.917 & 0.1785 \\
\code{bam} FE & 0.942 & 0.911 & 0.932 & 0.1814 \\
\code{bam} RE & 0.924 & 0.877 & 0.894 & 0.1779 \\
\midrule
\multicolumn{5}{l}{\textit{Panel B: $n=200$, $T=8$}} \\
\midrule
\code{bam} FD & 0.942 & 0.925 & 0.941 & 0.1294 \\
\code{bam} FE & 0.942 & 0.918 & 0.933 & 0.1386 \\
\code{bam} RE & 0.934 & 0.907 & 0.903 & 0.1376 \\
\midrule
\multicolumn{5}{l}{\textit{Panel C: $n=500$, $T=4$}} \\
\midrule
\code{bam} FD & 0.944 & 0.934 & 0.924 & 0.1235 \\
\code{bam} FE & 0.947 & 0.937 & 0.944 & 0.1252 \\
\code{bam} RE & 0.921 & 0.883 & 0.889 & 0.1230 \\
\bottomrule
\end{tabular}
\begin{minipage}{0.92\textwidth}
\footnotesize \tablenote\ Entries report coverage of penalty-adjusted confidence bands for the centered nonparametric component in the partially linear model. ``Avg. width'' is the average pointwise interval width.
\end{minipage}
\end{table}

\begin{table}[H]
\centering
\caption{Penalty-adjusted coverage results for centered $g$: nonparametric model}
\label{tab:g_coverage_np_penalty}
\begin{tabular}{lrrrr}
\toprule
Method & Avg. pointwise & Min. pointwise & Uniform & Avg. width \\
\midrule
\multicolumn{5}{l}{\textit{Panel A: $n=200$, $T=4$}} \\
\midrule
\code{bam} FD & 0.939 & 0.924 & 0.914 & 0.1769 \\
\code{bam} FE & 0.942 & 0.919 & 0.925 & 0.1811 \\
\code{bam} RE & 0.925 & 0.894 & 0.884 & 0.1775 \\
\midrule
\multicolumn{5}{l}{\textit{Panel B: $n=200$, $T=8$}} \\
\midrule
\code{bam} FD & 0.945 & 0.931 & 0.939 & 0.1279 \\
\code{bam} FE & 0.943 & 0.930 & 0.938 & 0.1389 \\
\code{bam} RE & 0.935 & 0.923 & 0.917 & 0.1378 \\
\midrule
\multicolumn{5}{l}{\textit{Panel C: $n=500$, $T=4$}} \\
\midrule
\code{bam} FD & 0.947 & 0.932 & 0.934 & 0.1221 \\
\code{bam} FE & 0.946 & 0.929 & 0.930 & 0.1248 \\
\code{bam} RE & 0.920 & 0.880 & 0.871 & 0.1226 \\
\bottomrule
\end{tabular}
\begin{minipage}{0.92\textwidth}
\footnotesize \tablenote\ Entries report coverage of penalty-adjusted confidence bands for the centered nonparametric component. ``Avg. width'' is the average pointwise interval width.
\end{minipage}
\end{table}

Several findings are worth highlighting. For inference on \(\beta\), the three \code{bam} methods yield similar finite-sample behavior. Under the null hypothesis, empirical rejection rates tend to lie between about 5\% and 6\%, close to the nominal 5\% level. Inference on \(g(\cdot)\) is more demanding. The \code{bam FE} and \code{bam FD} estimators tend to have coverage probabilities in the 92\% to 94\% range, while the penalized-unit-effect specification tends to have lower coverage, sometimes falling below 90\%. This pattern is consistent with the shrinkage interpretation of \code{s(unit, bs = "re")}: when unit effects are correlated with regressors, shrinking them toward zero can introduce bias that the confidence bands do not account for.

The modest deviations from nominal coverage are consistent with the broader literature on cluster-robust inference. \citet{imbens_kolesar_2016} and \citet{pustejovsky_tipton_2018} show that conventional heteroskedasticity- and cluster-robust procedures can have finite-sample distortions in various settings, motivating degrees-of-freedom and leverage-based corrections. \citet{cameron_miller_2015} and \citet{mackinnon_nielsen_webb_2023} emphasize that cluster-robust inference is asymptotic in the number of clusters, so modest over-rejection can remain in finite samples even when the number of clusters is not especially small. The penalized-spline setting adds a further source of finite-sample distortion because the covariance calculation treats the estimated smoothing parameter as fixed, ignoring the additional variability from its selection. 

Another possible reason for deviations from nominal coverage is smoothing bias in nonparametric series and spline estimators. Penalized series and spline estimators can have bias that is non-negligible relative to the standard error, and confidence bands built from the sandwich covariance matrix do not correct for this bias. The REML/fREML criterion targets prediction performance rather than coverage-optimal undersmoothing or bias correction, so the resulting bands can be slightly too narrow even when the variance estimate is accurate \citep{marra_wood_2012}. This effect is most directly relevant to inference on \(g(\cdot)\) rather than to tests of finite-dimensional parameters, which is consistent with the presented simulation evidence.

A wild cluster bootstrap that refits the \code{bam} model in each draw could improve finite-sample properties and is left for future work.

\section{Conclusion}\label{sec:conclusion}

This paper provides a practical guide to estimating semiparametric and nonparametric panel data models with fixed effects using the \code{mgcv} package in R. The proposed approach connects the penalized sieve and spline literature in theoretical econometrics with \mbox{REML/fREML} implementations of penalized splines in statistics. The fixed-effects estimators use unit indicators or first differencing, while penalized unit effects provide a regularized alternative with a distinct shrinkage interpretation. The same syntax can be applied to partially linear, varying-coefficient, fully nonparametric, and related specifications.

The focus is implementation. I provide \code{mgcv::bam} commands for estimation and show how to construct penalty-adjusted cluster-robust standard errors and confidence bands from the fitted model object. Monte Carlo simulations with correlated unit effects, heteroskedasticity, and serially correlated errors show good finite-sample performance in the one-dimensional smooth designs studied here: the estimators adapt reasonably well to the smoothness of the unknown function, tests for finite-dimensional parameters have empirical size close to nominal, and confidence bands for the centered nonparametric component have accurate coverage.

\bibliography{references}

\newpage
\begin{appendices}

\renewcommand{\thetable}{A\arabic{table}}
\renewcommand{\theHtable}{appendix.\arabic{table}}
\setcounter{table}{0}

\section{Additional Simulation Tables}

\begin{table}[H]
\centering
\footnotesize
\setlength{\tabcolsep}{4pt}
\renewcommand{\arraystretch}{0.92}
\caption{Classic coverage results for centered $g$}
\label{tab:g_coverage_classic_appendix}
\begin{tabular}{llrrrr}
\toprule
Model & Method & Avg. pointwise & Min. pointwise & Uniform & Avg. width \\
\midrule
\multicolumn{6}{l}{\textit{Panel A: $n=200$, $T=4$}} \\
\midrule
PL & \code{bam} FD & 0.996 & 0.982 & 1.000 & 0.2897 \\
PL & \code{bam} FE & 0.997 & 0.985 & 1.000 & 0.3018 \\
PL & \code{bam} RE & 0.996 & 0.980 & 1.000 & 0.3020 \\
\midrule
NP & \code{bam} FD & 0.996 & 0.984 & 1.000 & 0.2866 \\
NP & \code{bam} FE & 0.997 & 0.989 & 1.000 & 0.3008 \\
NP & \code{bam} RE & 0.995 & 0.982 & 0.999 & 0.3010 \\
\midrule
\multicolumn{6}{l}{\textit{Panel B: $n=200$, $T=8$}} \\
\midrule
PL & \code{bam} FD & 0.995 & 0.986 & 1.000 & 0.1923 \\
PL & \code{bam} FE & 0.996 & 0.990 & 0.999 & 0.2157 \\
PL & \code{bam} RE & 0.995 & 0.986 & 0.999 & 0.2157 \\
\midrule
NP & \code{bam} FD & 0.995 & 0.988 & 0.998 & 0.1897 \\
NP & \code{bam} FE & 0.996 & 0.989 & 1.000 & 0.2162 \\
NP & \code{bam} RE & 0.995 & 0.984 & 1.000 & 0.2162 \\
\midrule
\multicolumn{6}{l}{\textit{Panel C: $n=500$, $T=4$}} \\
\midrule
PL & \code{bam} FD & 0.994 & 0.986 & 1.000 & 0.1823 \\
PL & \code{bam} FE & 0.996 & 0.991 & 0.998 & 0.1898 \\
PL & \code{bam} RE & 0.992 & 0.973 & 0.997 & 0.1899 \\
\midrule
NP & \code{bam} FD & 0.994 & 0.989 & 1.000 & 0.1801 \\
NP & \code{bam} FE & 0.995 & 0.989 & 0.999 & 0.1893 \\
NP & \code{bam} RE & 0.991 & 0.966 & 0.998 & 0.1894 \\
\bottomrule
\end{tabular}
\begin{minipage}{0.92\textwidth}
\footnotesize \tablenote\ Entries report coverage for the series-style confidence bands that ignore the penalty. NP denotes the fully nonparametric model and PL denotes the partially linear model. ``Avg. width'' is the average pointwise interval width.
\end{minipage}
\end{table}

\end{appendices}

\end{document}